%% file: SiPMformulae-v1.tex
\documentclass[a4paper,english]{elsarticle}
\usepackage[T1]{fontenc}
\usepackage[latin9]{inputenc}
\usepackage{varioref}
\usepackage{units}
\usepackage{amsmath}
\usepackage{amssymb}
\usepackage{graphicx}
\usepackage{esint}
\usepackage{caption}
\usepackage{subcaption}

\makeatletter

\pdfpageheight\paperheight
\pdfpagewidth\paperwidth

\journal{Elsevier}
\usepackage{upgreek}
\usepackage[bookmarks,bookmarksopen,bookmarksdepth=6]{hyperref}
\usepackage{cases}
\usepackage{url}
\usepackage{lineno}
\usepackage{a4wide}

\makeatother

\usepackage{babel}
\begin{document}

\title{Characterisation of highly radiation-damaged SiPMs using current measurements}

\author[]{E.~Garutti}
\author[]{R.~Klanner\corref{cor1}}
\author[]{D.~Lomidze}
\author[]{J.~Schwandt}
\author[]{and M.~Zvolsky}
\cortext[cor1]{Corresponding author. Email address: Robert.Klanner@desy.de,
 Tel.: +49 40 8998 2558}
\address{Institute for Experimental Physics, University of Hamburg,
 \\Luruper Chaussee 147, D\,22761, Hamburg, Germany.}


\begin{abstract}
 The characterisation of radiation-damaged SiPMs is a major challenge, when the average time between dark counts approaches, or even exceeds, the signal decay time.
 In this note a collection of formulae is presented, which have been developed and used for the analysis of current measurements for SiPMs in the dark and illuminated by an LED, before and after hadron irradiation.
 It is shown, how parameters like the breakdown voltage, the quenching resistance, the dark-count rate, the reduction of the photo-detection efficiency due to dark counts and the Geiger discharge probability can be estimated from current-voltage measurements.
 The only additional SiPM parameters needed are the pixel capacitance, the number of pixels and the correlated noise.
 Central to the method is the concept of the pixel occupancy, the probability of a Geiger discharge in a single pixel during a given time interval, for which the decay time of the SiPM signal has been assumed. 
 As an illustration the formulae are used to characterise a KETEK SiPM before and after irradiation by a fluence of $5 \times 10^{13}$\,cm$^{-2}$ of reactor neutrons for temperatures of $- 30\,^\circ$C and $+ 20\,^\circ$C, where dark-count rates exceeding $10^{11}$\,Hz are observed.



\end{abstract}


\maketitle
\tableofcontents
 \pagenumbering{arabic}


 \section{Introduction}
 \label{sect:Introduction}
 Radiation damage by hadrons is one of the major limitations for the use of silicon photomultipliers (SiPM) at high luminosity accelerators and in space research.
  Accordingly, the investigation  of SiPM radiation damage and the improvement of their radiation hardness is a major research topic\,\cite{Musienko:2009,Quiang:2012,Heering:2016}.
 In this paper we develop methods to characterise SiPMs using current voltage measurements, which can also be applied at the high dark-count rates, $DCR$, where methods developed so far have difficulties.
 The data required are:
 \begin{enumerate}
   \item $I_{dark} (V_{for})$, the dark current measured for forward bias,
   \item $I_{dark} (V_{rev})$, the dark current measured for reverse bias, and
   \item $I_{dark+light} (V_{rev})$, the current measured with the SiPM illuminated by a DC light source.
 \end{enumerate}
 From 2. and 3., we obtain the additional current from the illumination
 \begin{equation}\label{equ:ILed}
   I_{light} = I_{dark+light} - I_{dark}.
 \end{equation}

 We note that for high DCR\,values, when the probability is significant that a dark count and a photon produce simultaneously an $eh$\,pair in the sensitive volume of the SiPM, $I_{light}$ depends both on the light intensity and on the DCR.

 For the analysis the following additional parameters, which cannot be obtained from current measurements, are required:
 \begin{enumerate}
   \item $N_{pix}$, the number of SiPM pixels,
   \item $C_{pix}$, the single pixel capacitance,
   \item $C_{q}$, the capacitance parallel to the quenching resistor, and
   \item $1 + CN$, the increase in SiPM signal due to correlated noise.
 \end{enumerate}
 The single pixel capacitance, $C_{pix}$ and $C_q$ have been obtained from the frequency dependency of the SiPM capacitance measured 0.5\,V below the breakdown voltage at 20\,$^\circ $C divided by $N_{pix}$,\,\cite{Chmill:2016}.
 For the formulae given, $C_q = 0$ is assumed.
 A finite value of $C_q$ significantly complicates several of the formulae.




 \section{Model and formulae}
  \label{sect:Formulae}

 \subsection{Quenching resistance $R_q$}
    \label{subsect:Rq}

 The quenching resistance, $R_q$, is obtained from $I_{dark} (V_{for})$ using
 \begin{equation}\label{equ:Rq}
   R_q = \Big(\frac{\textrm{d}I_{dark}} {\textrm{d}V_{for}}\Big)^{-1}.
 \end{equation}
 It has been noticed that $R_q(V_{for})$  has a finite slope even at voltages as high as 2\,V, and as noted in \cite{Chmill:2016}, the $R_q$\,value obtained from the frequency dependence of the capacitance 0.5\,V below the breakdown voltage is more reliable.

 \subsection{Rate of converting photons $R_\gamma$}
    \label{subsect:Rgamma}

 In this and in the following sections, we use $V$ for the reverse voltage, $V_{rev}$.
 The rate of photons which generate $eh$\,pairs in the sensitive volume of the SiPM, $R_\gamma $, is estimated from $I_{light}$ using
 \begin{equation}\label{equ:Rgamma}
   R_\gamma = \frac {I_{light}(V_{G=1})} {q_0},
 \end{equation}
 where a value of the reverse voltage $V_{G=1}$ has to be chosen, at which the SiPM gain $G = 1$.
 In the data analysis it should be checked that $I_{light}$ is constant in the region of $V_{G=1}$.

 We note that the assumptions that the $R_\gamma $ does not change with $V$ between $V_{G=1}$ and the SiPM operating voltage, is far from trivial.
 An increase in $V$ increases the depletion region of the $pn$ junction, which can increase the efficiency of an $eh$\,pair generated by light in the non-depleted region to reach the high electric-field region.
 In addition, radiation damage increases the silicon resistivity, and not all free charge carriers may be transported.
 However, as shown in Refs.\,\cite{Wiederspan:2017, Scharf:2017}, this effect can probably be ignored for the high fields present in SiPMs.

  \subsection{Breakdown voltage $V_{bd}$,  excess voltage $V_{ex}$, and gain $G$}
    \label{subsect:Vbd}

 For the determination of the breakdown voltage, $V_{bd}$, a method using the Inverse Logarithmic Derivative,
 $ILD = 1/\big(\frac{\textrm{d}\ln(I)} {\textrm{d}V}\big)$
 is recommended.
 A straight-forward way is to determine $V_{bd}$ as the voltage at which $ILD$ has its minimum\,\cite{Chmill:2016, Xu:2014}.
 The minimum can be obtained by a parabolic interpolation using the three $ILD$\,values around the minimum.
 Another way is to fit the rising part of $ILD$ with a first or second order polynomial and determine $V_{bd}$ as the voltage at which the polynomial crosses the $V$\,axis.
 We found that the results are quite similar and use the \emph{ILD-minimum method}.
 In the analysis it should be checked, that using $I_{dark}$, $I_{dark+light}$ and $I_{light}$ give compatible results for $V_{bd}$.

 The excess voltage is defined as
 \begin{equation}\label{equ:Vex}
   V_{ex} = V - V_{bd}.
 \end{equation}
 Values $V_{ex} > 0$ correspond to the SiPM operating range as photo detector.

 For the gain, G, the following, approximate relation is used
 \begin{equation}\label{equ:Gain}
   G = \frac{C_{pix} \cdot V_{ex}} {q_0}.
 \end{equation}
 A more accurate relation is:
  \begin{equation}\label{equ:GGain}
   G = \frac{(C_{pix} + C_q) \cdot (V - V_{to})} {q_0},
 \end{equation}
 where $C_q$ is an additional capacitance in parallel to $R_q$, and $V_{to}$ the voltage at which the Geiger discharge \emph{turns off}, as the current flowing through the pixel is too low to maintain a discharge.
 An additional capacitance $C_q$ results in a fast initial pulse, which is implemented in SiPMs with larger pixels to improve the time resolution.
 As far as we know, there is no fundamental reason why $V_{to} = V_{bd}$.
 In \cite{Chmill:2016} it has been shown that for the KETEK SiPM  with a pixel size of 15\,$\upmu $m
 $V_{bd} - V_{to} \approx 1$\,V, whereas for similar KETEK SiPMs with pixel sizes of 25\,$\upmu $m, 50\,$\upmu $m and 100\,$\upmu $m, $V_{to} \approx V_{bd}$ has been found.
 In the following, we will make the assumption of Eq.\,\ref{equ:Gain} for $G$.
 Using the relation Eq.\,\ref{equ:GGain} instead of Eq.\,\ref{equ:Gain} is straight-forward; this however is not the case for a precise determination of $C_q$\,\cite{Xu:2014} and of $V_{to}$\,\cite{Chmill:2016}.

 \subsection{Model for SiPM, pixel occupancy and photo-detection efficiency}
   \label{subsect:Models}

 For the relation between the dark current, $I_{dark}$, and the dark-count rate, $DCR$, the Gain, $G$, and the correlated noise, $CN$, we assume
 \begin{equation}\label{equ:Idark}
   I_{dark} ^{model} = q_0 \cdot DCR \cdot G \cdot (1+CN)
            = q_0 \cdot G \cdot \frac{N_{pix} \cdot \eta _{DC}} {\Delta t}.
 \end{equation}
 On the right-hand side, we have introduced the single-pixel occupancy, $\eta _{DC}$, which denotes the probability that a pixel, because of Dark Counts ($DC$s), is busy during the signal time $\Delta t$.
 The pixel occupancy includes the effects of the $CN$.
 From Eq.\,\ref{equ:Idark} we obtain $DCR \cdot (1 + CN) = N_{pix} \cdot \eta_ {DC}/\Delta t$.

 For the SiPM current with illumination in the presence of dark current, $I_{dark+light}$, we assume
  \begin{equation}\label{equ:IdarkLED}
   I_{dark+light}^{model} = q_0 \cdot G \cdot \frac{N_{pix} \cdot \eta _{DC+light}} {\Delta t},
 \end{equation}
 where $\eta _{DC+light}$  is the single-pixel occupancy due to $DC$s and illumination.

 For the SiPM current with illumination in the absence of a dark current, $I_{0\,dark+light}$, we assume
  \begin{equation}\label{equ:IdarkLED0}
   I_{0\,dark+light}^{model} = q_0 \cdot R_\gamma \cdot G \cdot (1 + CN) \cdot p_{Geiger}
   = q_0 \cdot G \cdot \frac{N_{pix} \cdot \eta _{\,light}} {\Delta t},
 \end{equation}
 where $R_\gamma $ is the rate of photons producing $eh$\,pairs in the sensitive volume of the SiPM, and $p_{Geiger}$ the probability of a photon to cause a Geiger discharge in the absence of pile-up due to the pixel occupancy by $DC$s.

 In all 3 cases the relation between current and occupancy assumed is
 \begin{equation}\label{equ:Ieta}
   I = \frac{q_0 \cdot N_{pix} \cdot G \cdot \eta } {\Delta t}.
 \end{equation}
 Assuming
 \begin{equation}\label{equ:tau}
   \Delta t = \tau \approx R_q \cdot C_{pix} \hspace{5mm}  \mathrm{and} \hspace{5mm}  G = C_{pix} \cdot V_{ex} / q_0,
 \end{equation}
 we obtain
 \begin{equation}\label{equ:eta}
   \eta  = \frac{I} {V_{ex}} \cdot \frac {R_q} {N_{pix}} = \frac{I} {I_{max}},
 \end{equation}
 where $\tau $ is the recharging time constant of the pixel, $R_q$ the quenching resistance, and $C_{pix}$ the pixel capacitance.
 The maximum current $I_{max} = N_{pix} \cdot V_{ex} / R_q $ corresponds to the situation for which the voltage drop over $R_q$ is $V_{ex}$, i.e. a continuous Geiger discharge.

 From $\eta $ we estimate $\mu $, the average number of $eh$\,pairs per time interval $\tau $, which would produce Geiger discharges, if there were no pile-up effects
  \begin{equation}\label{equ:mu}
   \mu  = -\ln(1 - \eta ).
 \end{equation}
 This relation is valid if the number of Geiger discharges is distributed according to a Poisson distribution, but also for a Generalised Poisson distribution\,\cite{Vinogradov:2012, Chmill:2017}, if the effect of the $CN$ is included in $\eta $, as is the case here.
 However, the relation is only valid if all pixels behave the same, in particular have the same $DCR$.
 In Ref.\,\cite{Engelmann:2016} the $DCR$ of individual pixels for non-irradiated KETEK SiPMs has been determined from the light produced by the Geiger discharges, and large pixel-to-pixel differences were observed.
 In Ref.\,\cite{Engelmann:2017} it is shown that these pixel-to-pixel differences even increase after irradiation with a fluence of $10^{10}$ thermal neutrons.

 As the average number of $eh$\,pairs produced by $DC$s, $\mu _{DC}$, and by the illumination, $\mu _{light}$, is additive, we obtain for the average number of $eh$\,pairs from the illumination
  \begin{equation}\label{equ:mu_LED}
   \mu _{light} = \mu _{DC+light} - \mu _{DC},
 \end{equation}
 and for the multiplicative factor to the photon detection efficiency ($pde$), $\varepsilon _{light}$, which accounts for the reduction of the $pde$ due to pixel occupancy
  \begin{equation}\label{equ:eps_light}
   \varepsilon _{light} = \frac {\eta _{DC+light} - \eta _{DC}} {\mu _{light}} =
   \frac {\eta _{DC+light} - \eta _{DC}} {\ln \Big(\frac{1 - \eta _{DC}} {1 - \eta _{DC+light}}  \Big)}.
 \end{equation}
 The current increase due to the illumination, which can be compared to $I_{light}$ defined in Eq.\,\ref{equ:ILed}, is given by
  \begin{equation}\label{equ:I_LEDmodel}
 I_{dark+light}^{model} - I_{dark}^{model}
 =  I_{0dark+light}^{model} \cdot \varepsilon _{light}
 = q_0 \cdot G \cdot (1 + CN) \cdot R_\gamma \cdot p_{Geiger} \cdot \varepsilon _{light}.
 \end{equation}
 As mentioned in Sect.\,\ref{sect:Introduction}, the additional current due to the illumination depends on $R_\gamma $ as well as on the $DCR$.
 Eq.\,\ref{equ:I_LEDmodel} can also be used to determine $p_{Geiger}$ using only current measurements and an estimate of $CN$.

 We note that for $(\eta _{DC+light} - \eta_{DC}) / ( 1 - \eta_{DC}) \ll 1$, which corresponds to a low light intensity, the series expansion of Eq.\,\ref{equ:eps_light} gives $\varepsilon _{light} \approx 1 - \eta_{DC}$.
 Thus the efficiency of the detection of photons is reduced by the probability that a Geiger discharge from a dark count already has occurred.

 \subsection{Normalised SiPM currents with light}
    \label{subsect:ILED}

 A straight-forward method of obtaining an idea of the effects of radiation damage or temperature on the SiPM performance is to compare  $I_{light}^{norm} = I_{light} / R_\gamma$ of a SiPM for different irradiation and measurement conditions\,\cite{Vignali:2016, Vignali:2017}.
 From Eq.\,\ref{equ:I_LEDmodel} we obtain
 \begin{equation}\label{equ:ILEDratio}
   \frac{I_{light}^{\Phi_2,T_2}/R_\gamma ^{\Phi_2,T_2}} {I_{light}^{\Phi_1,T_1}/R_\gamma ^{\Phi_1,T_1}}
   = \frac{G^{^{\Phi_2,T_2}} \cdot (1 + CN^{\Phi_2,T_2}) \cdot p_{Geiger}^{\Phi_2,T_2} \cdot \varepsilon _{light}^{\Phi_2,T_2} }
          {G^{^{\Phi_1,T_1}} \cdot (1 + CN^{\Phi_1,T_1}) \cdot p_{Geiger}^{\Phi_1,T_1} \cdot \varepsilon _{light}^{\Phi_1,T_1}},
 \end{equation}
 were the $\Phi_i $ denote the irradiation fluences and $T_i$ the temperatures of two measurements.
 If $V_{bd}$ is the only SiPM parameter which changes and the values of $\varepsilon _{light}$ are the same, the $I_{light} ^{norm}$ ratio as function of excess voltage, will be 1.
 A deviation from 1 indicates a change of at least one of the SiPM parameters or of the occupancy.
 A systematic fluctuation of the ratio in the region of $V_{bd}$, where the shape of $I_{light}$ varies rapidly, indicates a mismatch of the assumed $V_{bd}$ values.

 This method can be used to investigate the fluence dependence of the SiPM performance at a fixed temperature.
 For a non-irradiated SiPM with low $DCR$ and illuminated well below saturation, $\varepsilon _{light} = 1$ can be assumed, and the $I_{light}^{norm}$\,ratio for the fluences $\Phi _1 $ to $\Phi _2 = 0$ as a function of $V_{ex}$ will show, if the SiPM performance has changed.
 In a similar way the ratio, after taking into account the change of $V_{bd}$ with temperature, can be used to investigate possible changes of the product $G \cdot (1 + CN) \cdot p_{Geiger} \cdot \varepsilon _{light}$ with temperature.

 \subsection{Dark-count rate $DCR$}
    \label{subsect:DCR}

 Using Eqs.\,\ref{equ:Idark} and \ref{equ:tau} one finds
 \begin{equation}\label{equ:DCR}
   DCR = \frac{N_{pix} \cdot \eta _{DC}} {R_q \cdot C_{pix}} \cdot \frac{1} {1 + CN}
       = \frac{I_{dark}} {C_{pix} \cdot V_{ex} \cdot (1 + CN)},
 \end{equation}
 which can be used to estimate the $DCR$.
 Another way of estimating the $DCR$ for high fluences is to assume that $G$ and $CN$, after taking into account a possible change of $V_{bd}$, do not depend on fluence.
 Then the $DCR$ scales with $I_{dark}$, and, if the $DCR$ at a low fluence $\Phi _1$ is known, the approximate $DCR$ at a high fluence $\Phi _2$ is given by:
 \begin{equation}\label{equ:DCRscaling}
   DCR_{\Phi_ 2} = \frac{I_{dark}^{\Phi _2}} {I_{dark}^{\Phi _1}} \cdot DCR_{\Phi_ 1}.
 \end{equation}
 This relation is useful, if the $DCR$ at a low fluence $\Phi _1$ has been determined from the pulse height spectrum measured in the dark.
 However, the assumption that $G$ and $CN$ are the same for both conditions has to be checked, which can be done using the ratio of normalised currents with illumination discussed in Sect.\,\ref{subsect:ILED}, or $C_{pix}$ determinations using capacitance--frequency measurements at voltages close to $V_{bd}$\,\cite{Xu:2014}.

 \section{Measurements and data analysis}
  \label{sect:Analysis}

 In this section we use some of the formulae presented in Sect.\,\ref{sect:Formulae} to characterise a KETEK SiPM with a $15\,\upmu$m pixel size before and after irradiation by reactor neutrons with a dose of $5 \times 10^{13}$\,cm$^{-2}$.
 The number of pixels $N_{pix} = 4384$, and in\,\cite{Chmill:2016} $C_{pix} = 18$\,fF and $C_q < 5$\,fF have been determined with capacitance-frequency measurements 0.5\,V below $V_{bd}$ at $20\,^\circ $C\,\cite{Chmill:2016}.
 The following current measurements, taken at $- 30\,^\circ $C and $ + 20\,^\circ $C
  \begin{enumerate}
   \item $I_{dark} (V_{for})$, the dark current measured for forward bias between 0 and 2\,V,
   \item $I_{dark} (V_{rev})$, the dark current measured for reverse bias, between 0 and 35\,V, and
   \item $I_{dark+light} (V_{rev})$, the current measured with the SiPM illuminated by a blue LED, between 0 and 40\,V.
 \end{enumerate}
 were analysed.
 For 3., measurements at two different LED light intensities, "low LED" and "high LED", were made.
 To check the quality of the measurements, the voltage was ramped up and down for $V_{for}$ and $V_{rev}$, and it was checked if the results agree.
 With the exception of the measurements at $- 30\,^\circ $C and low LED, the agreement was within 1\,\%.
 For the $- 30\,^\circ $C data, discrepancies at the 30\,\%\,level where observed.
 Unfortunately the measurements could not be repeated, as this SiPM stopped working.
 We note that we do not have a complete data set for the same SiPM before and after irradiation, and the measurements for the non-irradiated and irradiated SiPM come from different samples.
 As a result, some of the differences of the SiPM parameters before and after irradiation, in particular the values of $R_q$ and $V_{bd}$, are ascribed to the different SiPMs.


 \subsection{Quenching resistance $R_q$}
  \label{sect:Rq-analysis}

 The quenching resistance, $R_q$, was determined from $I_{dark}$ measured at $V_{for}$ at 1.6 and 1.8\,V using Eq.\,\ref{equ:Rq}.
 We could not use the data at 2\,V, because the current exceeded the current limit of voltage source for some measurements.
 The results are presented in Table\,\ref{tab:RqVbd}.
 We ascribe most of the increase of $R_q$ with fluence to irradiation effects:
 Measurements at $+20 \,^\circ $C of the same SiPM before and after irradiation show an increase of $R_q$ by $\approx 40 $\,\% after a fluence of $\Phi = 5 \times 10^{13}$\,cm$^{-2}$.
 In addition, sample to sample differences of up to $\pm \,30$\,\% have been observed.
 A decrease of $R_q$ with temperature is expected for a poly-Si resistor, due to increase of the intrinsic charge-carrier concentration.
 For the ratio $R_q (-30 \,^\circ $C)/$R_q (+20 \,^\circ $C) a value of 0.645 is observed for both non-irradiated and irradiated SiPM.
 We also note that the values of $R_q$ determined using $C-V$ measurements $\approx 0.5$\,V below $V_{bd}$\,\cite{Chmill:2016}, are about 10\,\% lower than the ones obtained from  $I-V_{for}$.
 As the value of $R_q$ obtained from the $I-V_{for}$\,measurements depends on the choice of the $V_{for}$\,interval, we consider the $C-V$ results to be more accurate.

\begin{table}  [!ht]
 \centering
  \footnotesize
 \begin{tabular}{c||c|c|c||c|c}
  \centering
    & $R_q$\,[k$\Omega$] & $R_q$\,[k$\Omega]$  & $\underline{\hspace{5mm} R_q(0)\hspace{5mm}}$ & $V_{bd}$\,[V] & $V_{bd}$\,[V] \\
   $T$\,[$^\circ $C]  & $\Phi = 0$\,cm$^{-2}$ & $\Phi = 5 \times 10^{13}$\,cm$^{-2}$ & $R_q(5 \times 10^{13})$ &$\Phi = 0$\,cm$^{-2}$ & $\Phi = 5 \times 10^{13}$\,cm$^{-2}$ \\
     \hline \hline
   $- 30$ & $811 \pm 40$ & $1313 \pm 65$ & $1.61 \pm 0.11$ & $26.298 \pm 0.010$ &$26.450 \pm 0.010$ \\
      \hline
   $+ 20$ & $519 \pm 26$  & $849 \pm 43$ & $1.63 \pm 0.13$ & $27.391 \pm 0.010$ & $27.557 \pm 0.010$ \\
    \hline \hline
 \end{tabular}
   \caption{Values of $R_q$ and $V_{bd}$ from the $I-V$\,measurements. The errors given present the uncertainties of the data and analysis, but do not include systematics.
   The measurements for the non-irradiated and irradiated SiPM were made on different samples.
   The differences for the two dose values in $R_q$ are ascribed mainly to the irradiation, and the ones in $V_{bd}$ to differences between the two SiPM samples and not to irradiation effects.
   \label{tab:RqVbd}}
\end{table}

\begin{figure}[!ht]
   \centering
   \begin{subfigure}[a]{0.5\textwidth}
    \includegraphics[width=\textwidth]{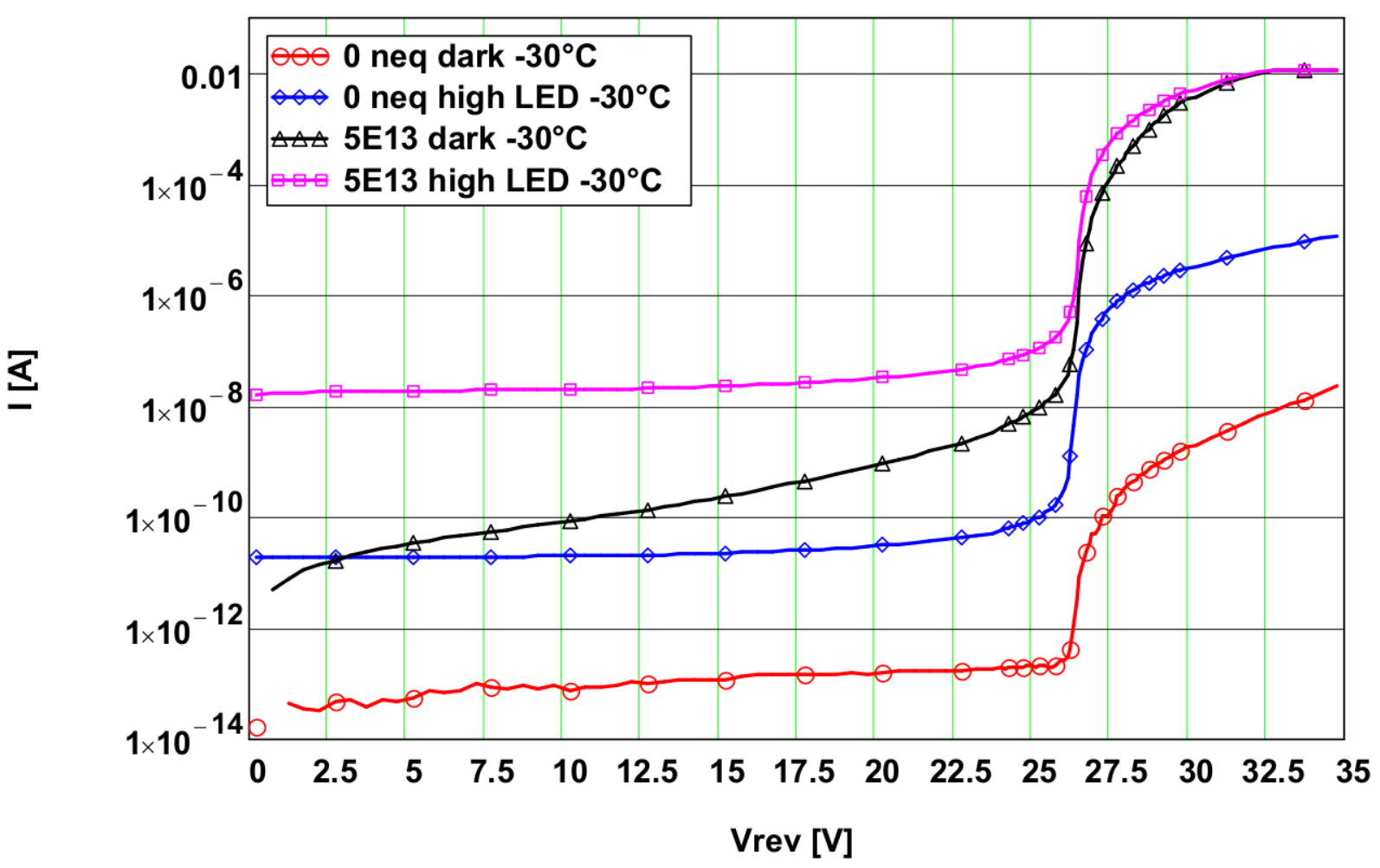}
    \caption{ }
   \end{subfigure}%
    ~
   \begin{subfigure}[a]{0.5\textwidth}
    \includegraphics[width=\textwidth]{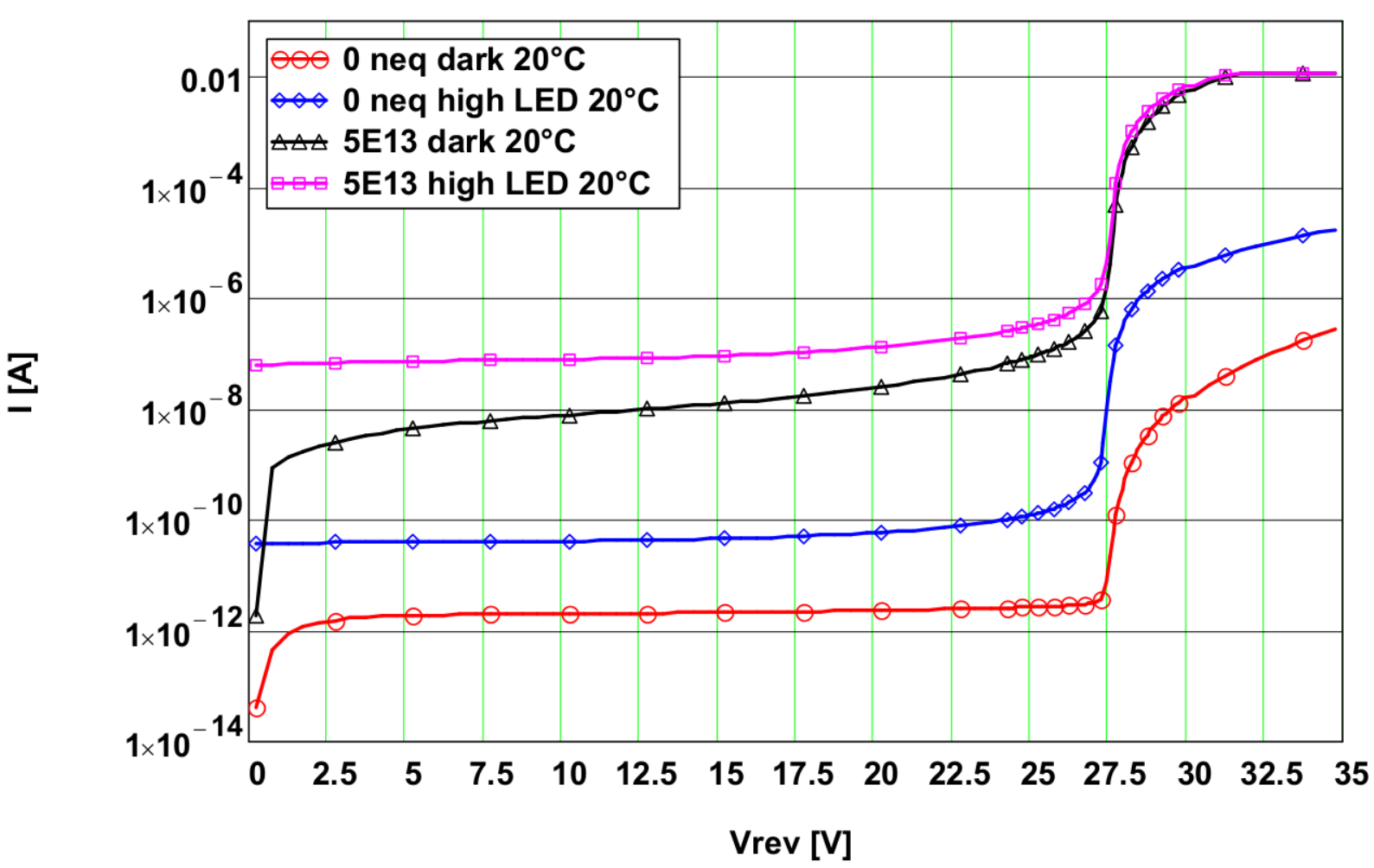}
    \caption{ }
   \end{subfigure}%
   \caption{ Comparison of the $I-V_{rev}$ measurements before and after neutron irradiation to $5 \times 10^{13}$\,cm$^{-2}$ with and without LED illumination at (a) $-30$\,$^\circ $C, and (b) $+20$\,$^\circ $C.
   The current limit of the voltage source is set to 12\,mA, which causes the current saturation at high voltages. }
  \label{fig:IV}
 \end{figure}

  \subsection{Breakdown voltage $V_{bd}$}
  \label{sect:Vbd-analysis}

 Fig.\,\ref{fig:IV} shows the $I-V_{rev}$ data.
 Below $V_{bd} \approx 27$\,V, the dark current, $I_{dark}$, for the non-irradiated SiPM is approximately constant and hardly shows an increase due to avalanche multiplication, when approaching $V_{bd}$.
 We conclude that for the non-irradiated sensor most of the $I_{dark}$ misses the amplification region and is not generated in the sensitive region of the SiPM.
 After irradiation, $I_{dark}$ increases by $3 - 4$ orders of magnitude, and a continuous increase of $I_{dark}$ with voltage as well as avalanche multiplication are observed, which indicates that most of $I_{dark}$ is generated in the sensitive region of the SiPM.
 Above the $V_{bd}$ and below the current limit of the voltage source of 12\,mA, $I_{dark}$ increases by about 6 orders of magnitude due to the irradiation, but the shapes of the $I-V$\,curves for the non-irradiated and irradiated SiPM are similar.

 The voltage dependencies of the currents with LED\,irradiation, $I_{dark+light}$, are very similar and appear to depend neither on irradiation fluence nor on temperature.
 $I_{dark+light}$ is essentially constant at low voltages, shows an increase due to avalanche multiplication when approaching $V_{bd}$, and, when passing and exceeding $V_{bd}$, the expected rapid increase due to Geiger discharges.

\begin{figure}[!ht]
   \centering
    \includegraphics[width=0.5\textwidth]{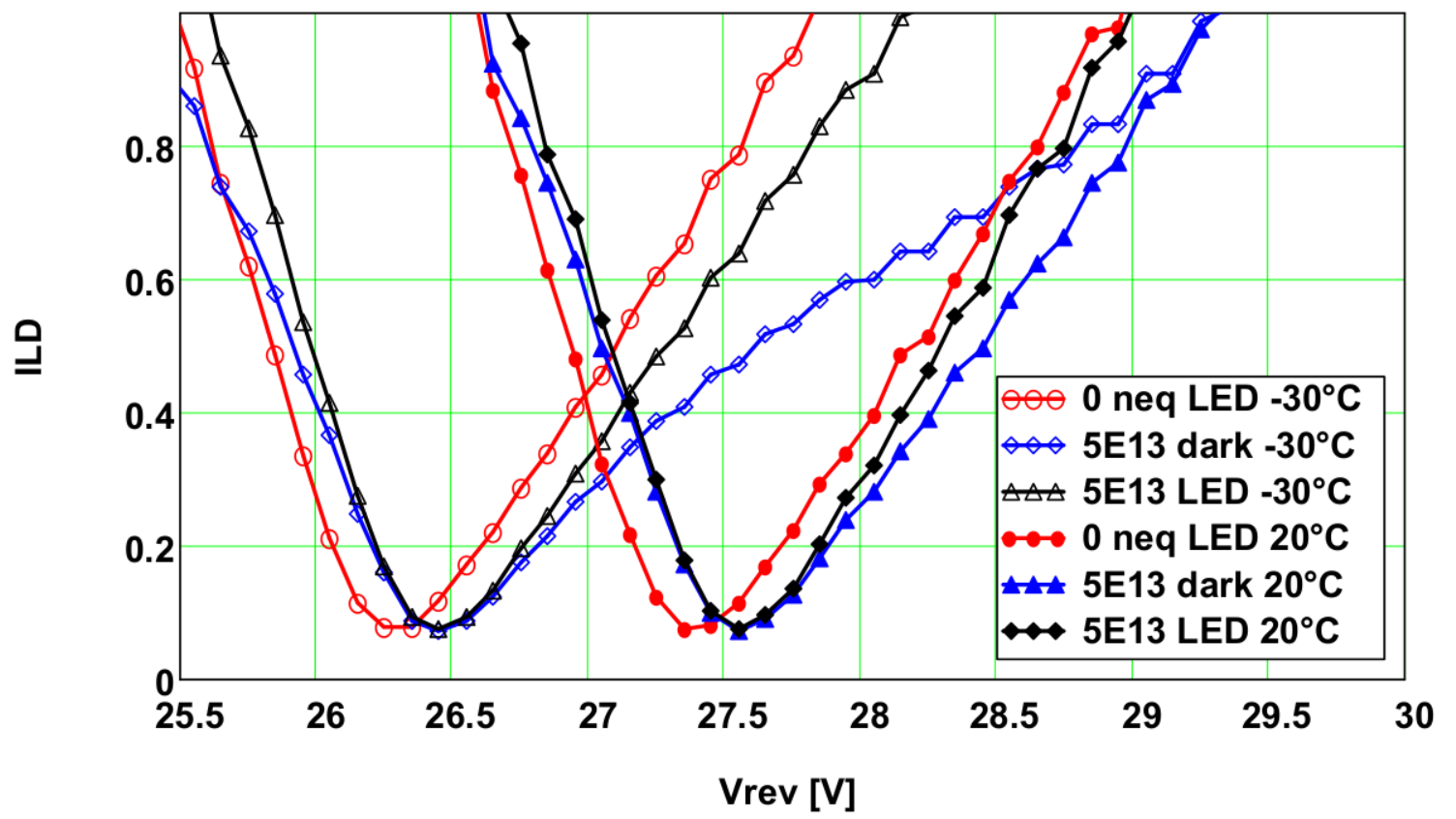}
   \caption{Inverse Logarithmic Derivative, $ILD$, for the $I - V$ data shown in Fig.\,\ref{fig:IV}. }
   \label{fig:ILD}
 \end{figure}

 From the $I-V$\, data the breakdown voltage, $V_{bd}$ has been determined using the \emph{minimum ILD} method discussed in Sect.\,\ref{subsect:Vbd}.
 Fig.\,\ref{fig:ILD} shows the $ILD$\,curves and Table\,\ref{tab:RqVbd} the $V_{bd}$\,values.
 The $ILD$\,curves from $I_{dark}$ of the non-irradiated SiPM are not shown, as $I_{dark}$ of the non-irradiated SiPM is low, resulting in big $ILD$\,fluctuations.

 The values of $V_{bd}$ obtained from $I_{dark}$ and $I_{dark+light}$ agree within their uncertainties.
 For the two temperatures, within the experimental uncertainties, the same difference ($152 \pm 14$\,mV at $-30 ^\circ $C and $166 \pm 14$\,mV at $+20 ^\circ $C) for the $V_{bd}$\,values for the non-irradiated and irradiated SiPM is observed.
 We ascribe this difference to the different SiPM samples investigated.
 In Ref.\,\cite{Vignali:2017}, using measurements at $ + 20\,^\circ $C it is shown that for the KETEK SiPM investigated, $V_{bd}$ does not change with irradiation up to a fluence of $5 \times 10^{13}$\,cm$^{-2}$.
 For both, the non-irradiated and the irradiated SiPM, the temperature dependence of $V_{bd}$ is 22.0\,mV/$^\circ $C.

 \subsection{Normalised SiPM currents with LED illumination}
  \label{sect:ILEDanalysis}

 From the measured currents with and without LED illumination, $I_{light}$ is determined using Eq.\,\ref{equ:ILed}, and with the help of Eq.\,\ref{equ:Rgamma}, $R_\gamma $, the rate of photons generating $eh$\,pairs in the sensitive region of the  SiPM.
 The results for $R_\gamma$ are shown in Table\,\ref{tab:Rgamma}.
 For $V_{G = 1} = 10$\,V has been chosen.
 Taking values of 5\,V or 15\,V changes the $R_\gamma$\,values by less than 5\,\%.

\begin{table}  [!ht]
 \centering
  \footnotesize
 \begin{tabular}{c||c|c||c|c}
  \centering
    & $R_\gamma $\,low LED [s$^{-1}]$ & $R_\gamma $\,high LED [s$^{-1}]$ & $R_\gamma $\,low LED [s$^{-1}]$ & $R_\gamma $\,high LED [s$^{-1}]$ \\
   $T$\,[$^\circ $C]  & $\Phi = 0$\,cm$^{-2}$ & $\Phi = 0$\,cm$^{-2}$ &$\Phi = 5 \times 10^{13}$\,cm$^{-2}$& $\Phi = 5 \times 10^{13}$\,cm$^{-2}$ \\
     \hline \hline
   $- 30$ & $6.2 \times 10^{7}$ & $12.5 \times 10^{7}$ & $6.9 \times 10^{10}$ & $20.3 \times 10^{10}$ \\
      \hline
   $+ 20$ & $12.5 \times 10^{7}$ & $24.8 \times 10^{7}$ & $12.5 \times 10^{10}$ & $44.5 \times 10^{10}$ \\
    \hline \hline
 \end{tabular}
   \caption{Values of $R_ \gamma$ from the $I_{light} - V$ data using Eq.\,\ref{equ:Rgamma} with $V_{G=1} = 10$\,V.
   \label{tab:Rgamma}}
\end{table}

\begin{figure}[!ht]
   \centering
   \begin{subfigure}[a]{0.5\textwidth}
    \includegraphics[width=\textwidth]{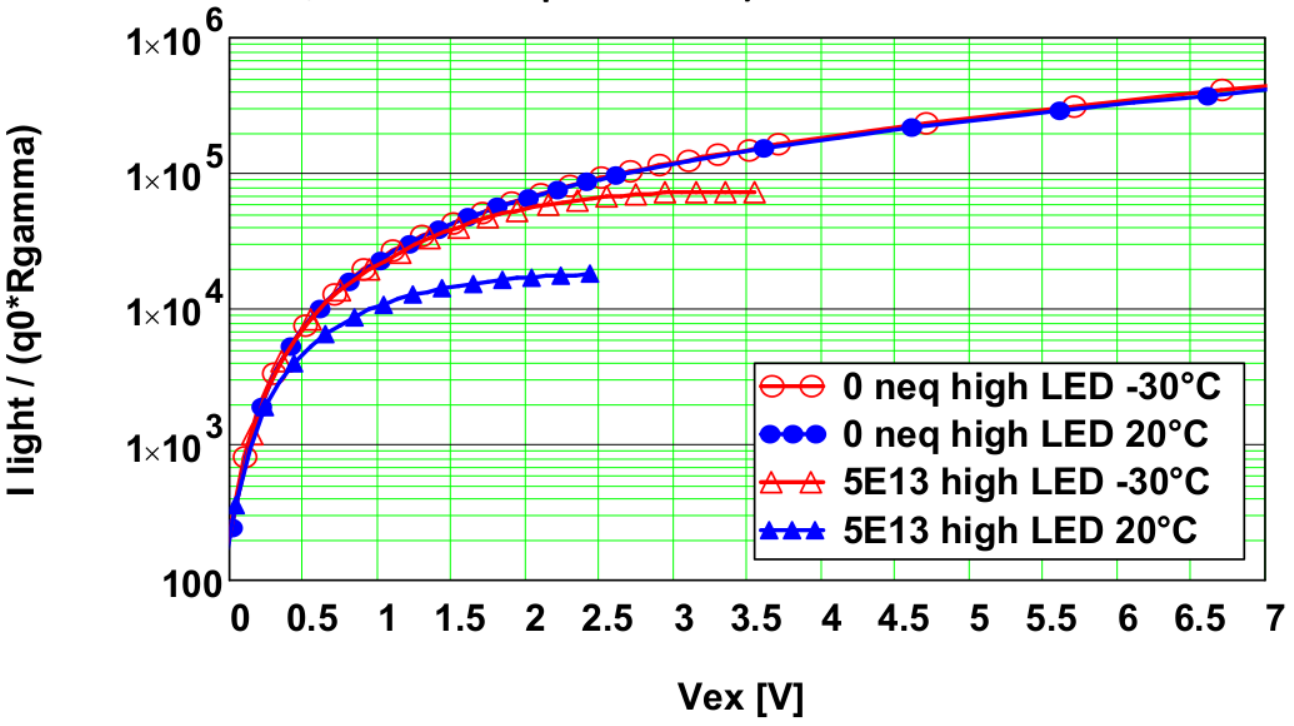}
    \caption{ }
   \end{subfigure}%
    ~
   \begin{subfigure}[a]{0.5\textwidth}
    \includegraphics[width=\textwidth]{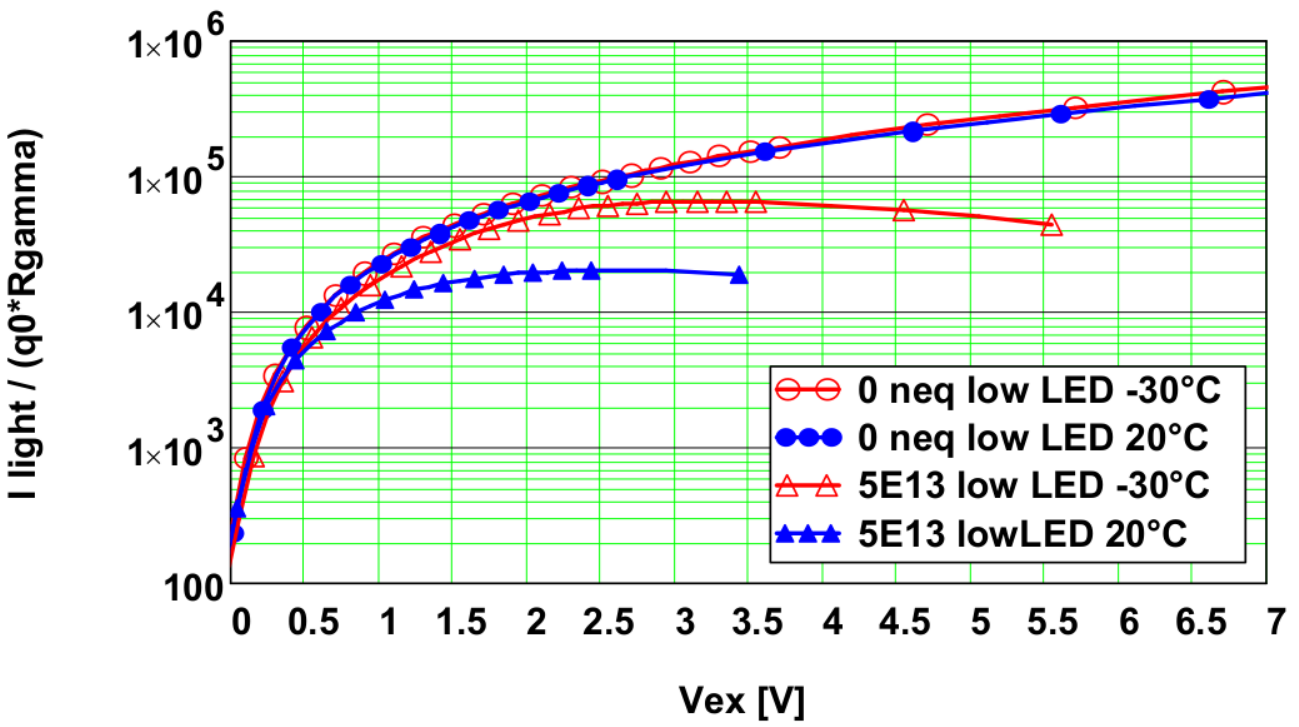}
    \caption{ }
   \end{subfigure}%
   \caption{Normalised $I_{light}$ as function of $V_{ex}$ for (a) "high LED", and (b) "low LED".
   According to Eq.\,\ref{equ:I_LEDmodel}, $I_{light}/(q_0 \cdot R_\gamma) = G \cdot (1 + CN) \cdot p_{Geiger} \cdot \varepsilon _{light}$.}
  \label{fig:ILEDRg}
 \end{figure}

 Fig.\,\ref{fig:ILEDRg} shows the normalised current with illumination, $I_{light} ^{norm} = I_{light} / (q_0 \cdot R_\gamma )$ as a function of $V_{ex}$.
 Up to $V_{ex} \approx 0.5\,V$, $I_{light} ^{norm}$ neither depends on temperature nor on fluence, and we conclude, as discussed in Sect.\,\ref{subsect:ILED}, that in this voltage range the SiPM performance is the same for the different measurement conditions.
 At $\approx 0.5$\,V the $\Phi = 5 \times 10^{13}$\,cm$^{-2}, T = 20\,^\circ $C, and at $\approx 2$\,V the $\Phi = 5 \times 10^{13}$\,cm$^{-2}, = -30\,^\circ $C\,curves start to deviate from the $\Phi = 0$\, results, and become constant at 2.5\,V and 3.5\,V, respectively.
 At these voltages the SiPM is no more a useful photo-detector, as the gain $G$, which is proportional to $V_{ex}$, is compensated by the loss in detection efficiency.
 For the $\Phi = 0$ data, the dependence of $I_{light} ^{norm}$ on $V_{ex}$ is approximately independent of temperature and light intensity.

\begin{figure}[!ht]
   \centering
   \begin{subfigure}[a]{0.5\textwidth}
    \includegraphics[width=\textwidth]{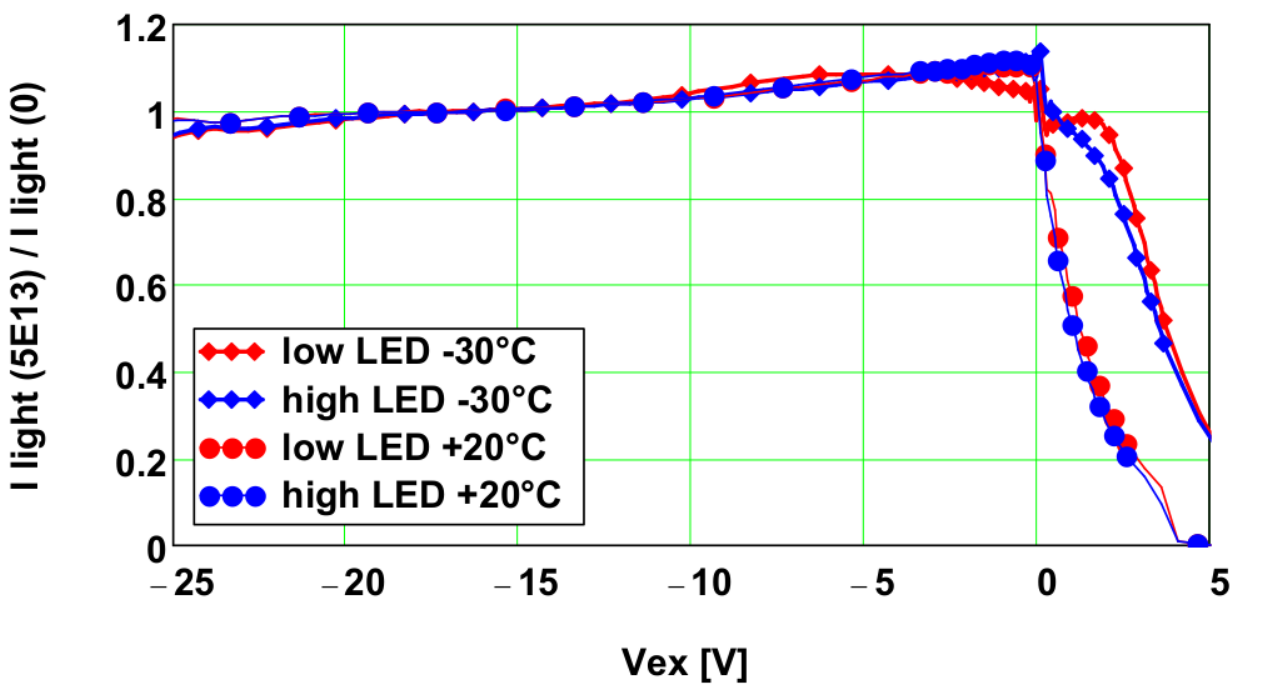}
    \caption{ }
   \end{subfigure}%
    ~
   \begin{subfigure}[a]{0.5\textwidth}
    \includegraphics[width=\textwidth]{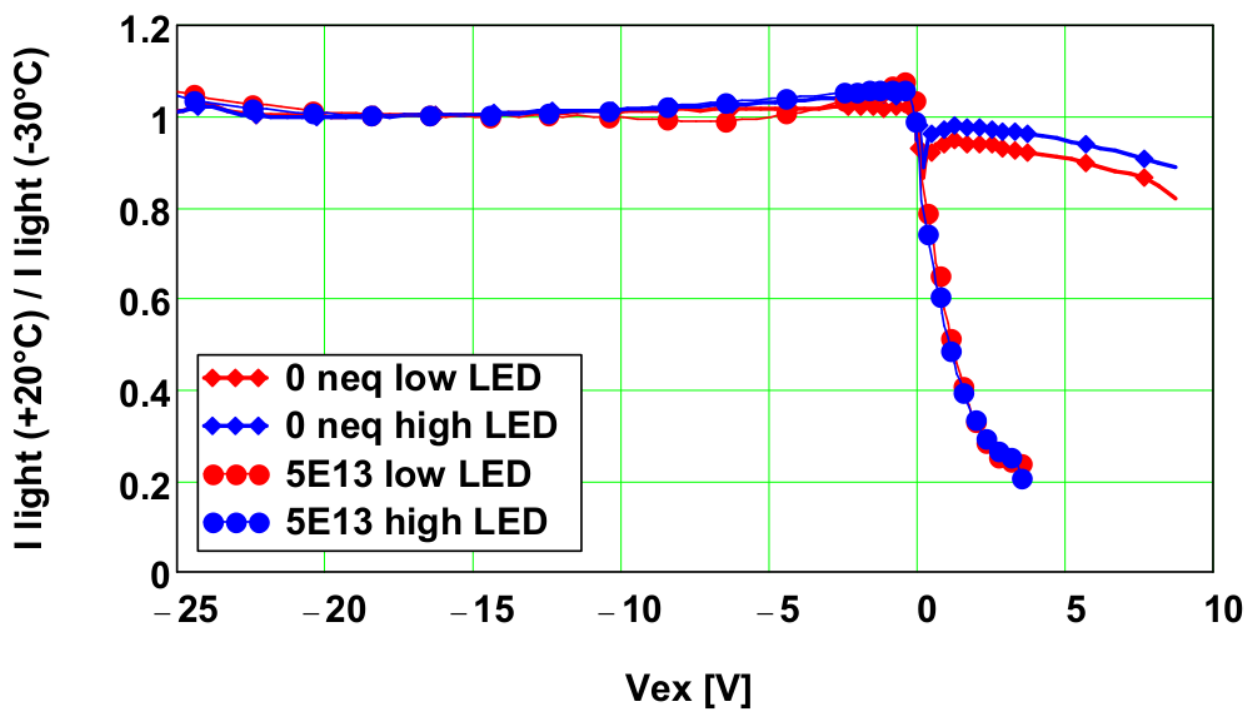}
    \caption{ }
   \end{subfigure}%
   \caption{Normalised current ratio with illumination (Eq.\,\ref{equ:ILEDratio}) as a function of $V_{ex}$.
   (a) $(I_{light}^{\,5\times 10^{13},T}/R_\gamma ^{\,5\times 10^{13},T})$/ $(I_{light}^{\,0,T}/R_\gamma ^{\,0,T})$ at $T = -30\,^\circ$C and $ +20\,^\circ$C, and
   (b) $(I_{light}^{\Phi, +20^\circ \mathrm{C}}/R_\gamma ^{\Phi, +20^\circ \mathrm{C}})/(I_{light}^{\Phi, -30^\circ \mathrm{C}}/R_\gamma ^{\Phi, -30^\circ \mathrm{C}})$ at $\Phi = 0$ and $ 5\times 10^{13}$\,cm$^{-2}$.}
  \label{fig:ILEDnorm}
 \end{figure}

 Fig.\,\ref{fig:ILEDnorm}\,a) shows the ratios of the normalised currents with illumination, defined in Eq.\,\ref{equ:ILEDratio}, for $\Phi _2 = 5 \times 10^{13}$\,cm$^{-2}$ to $\Phi _1 = 0$ as a function of $V_{ex}$ of the $ - 30\,^\circ$C and of $ + 20\,^\circ$C\,data for the low and the high LED intensity.
 As the $V_{bd}$\,values are different for the $\Phi _1$ and the $\Phi _2$\,data, a cubic spline interpolation has been used to calculate the values of $I_{light} ^{\Phi _2, \, T}$ at the voltages of the $\Phi _1$\,measurements.
 Per definition, the ratio at $V_{ex} \approx -17$\,V, which corresponds to $V_{rev} = V_{G=1} = 10$\,V, is 1.
 The ratio rises up to $\approx 1.1$ at the breakdown voltage, $V_{ex} = 0$, and then drops to zero due to the reduction of the photo-detection efficiency.
 The rise for $V_{ex} < 0 $ may be evidence for differences in the increase of the multiplication gain with voltage below $V_{bd}$.
 As will be shown in Sect,\,\ref{sect:eta-analysis}, the main reason for the decrease for $V_{ex} > 0 $ is the increase in pixel occupancy by dark counts and photons.
 As expected from the pixel occupancy, the decrease at $+20$\,$^\circ $C is faster than at $-30$\,$^\circ $C, and also faster for the high than for the low LED intensity.
 We note a difference in shape of the $-30$\,$^\circ $C low LED curve compared to the other 3 curves, which we ascribe to the measurement problem discussed at the beginning of Sect.\,\ref{sect:Analysis}.


 Fig.\,\ref{fig:ILEDnorm}\,b) shows the ratios of the normalised currents with illumination for $T_2 = + 20\,^\circ$C to $T_1 = - 30\,^\circ$C at $\Phi = 0$ and $ 5\times 10^{13}$\,cm$^{-2}$ for the low and high LED intensities.
 For the non-irradiated SiPM we note a shift of the ratio by about $+ 3$\,\% between the high- and low-LED data for voltages around and above $V_{ex} = 0$, which we do not understand and which may indicate a measurement problem.
 At $V_{ex} = 0$ the ratios drop by $\approx 7$\,\%, with a further decrease by about the same amount up to $V_{ex} = 8$\,V, the maximum value of the measurements.
 The results suggest that the value of $p_{Geiger} \cdot (1 + CN)$ decreases with increasing temperature.
 However, further studies, in particular a comparison to pulse-height measurements as a function of temperature, are required to verify this conclusion.

 For the irradiated SiPM the current ratio drops from 1 to $\approx 0.2$ between $V_{ex} = 0$ and 3\,V, where the measurements at $+ 20 \, ^\circ$C reach the current limit.
 As will be shown in Sect.\,\ref{sect:eta-analysis} this rapid decrease of the ratio is due to the increase of $DCR$ between $ 30 \, ^\circ$C and $+ 20 \, ^\circ$C.

  \subsection{Pixel occupancy $\eta $, photo-detection efficiency and Geiger-discharge probability}
  \label{sect:eta-analysis}

\begin{figure}[!ht]
   \centering
   \begin{subfigure}[a]{0.5\textwidth}
    \includegraphics[width=\textwidth]{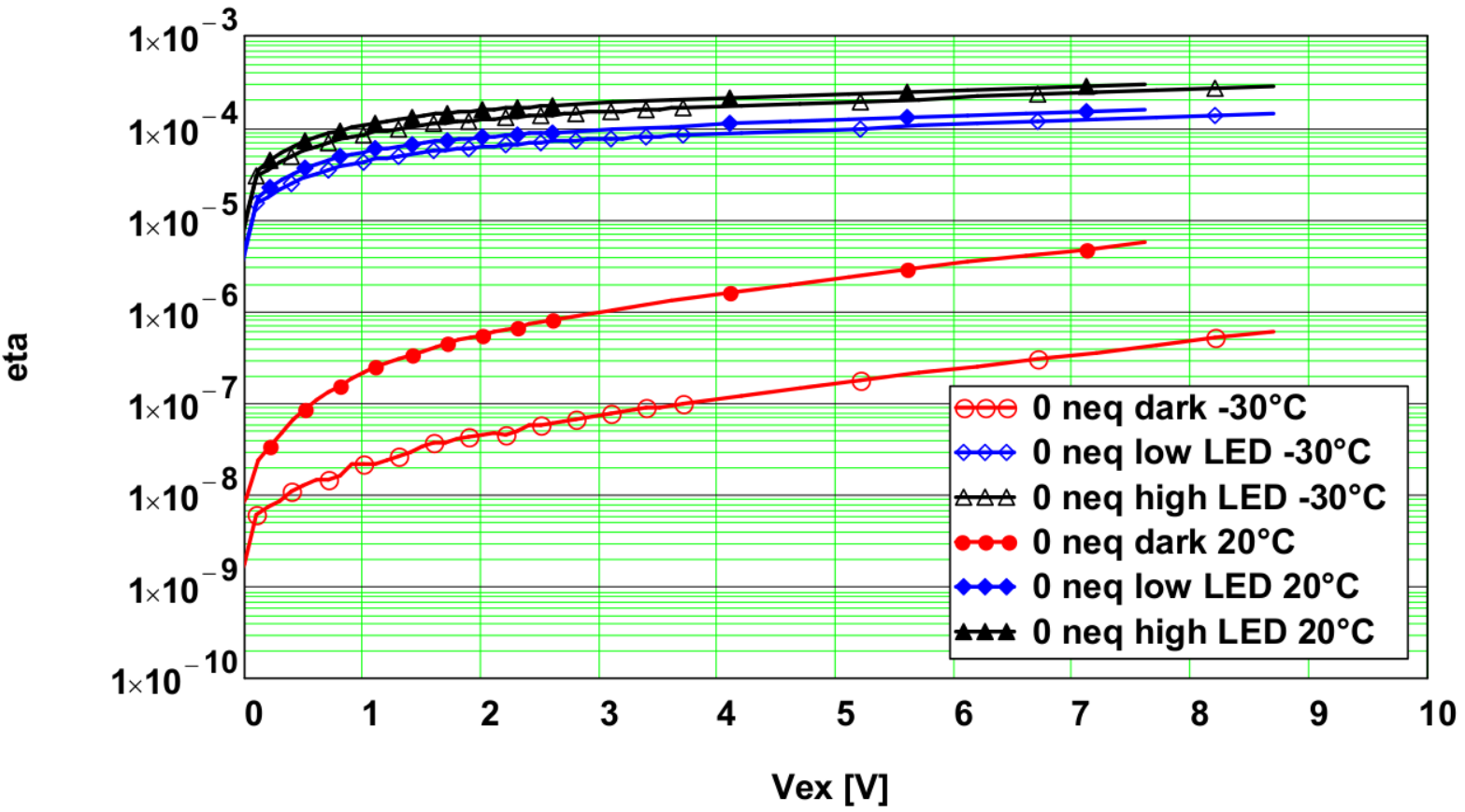}
    \caption{ }
   \end{subfigure}%
    ~
   \begin{subfigure}[a]{0.5\textwidth}
    \includegraphics[width=\textwidth]{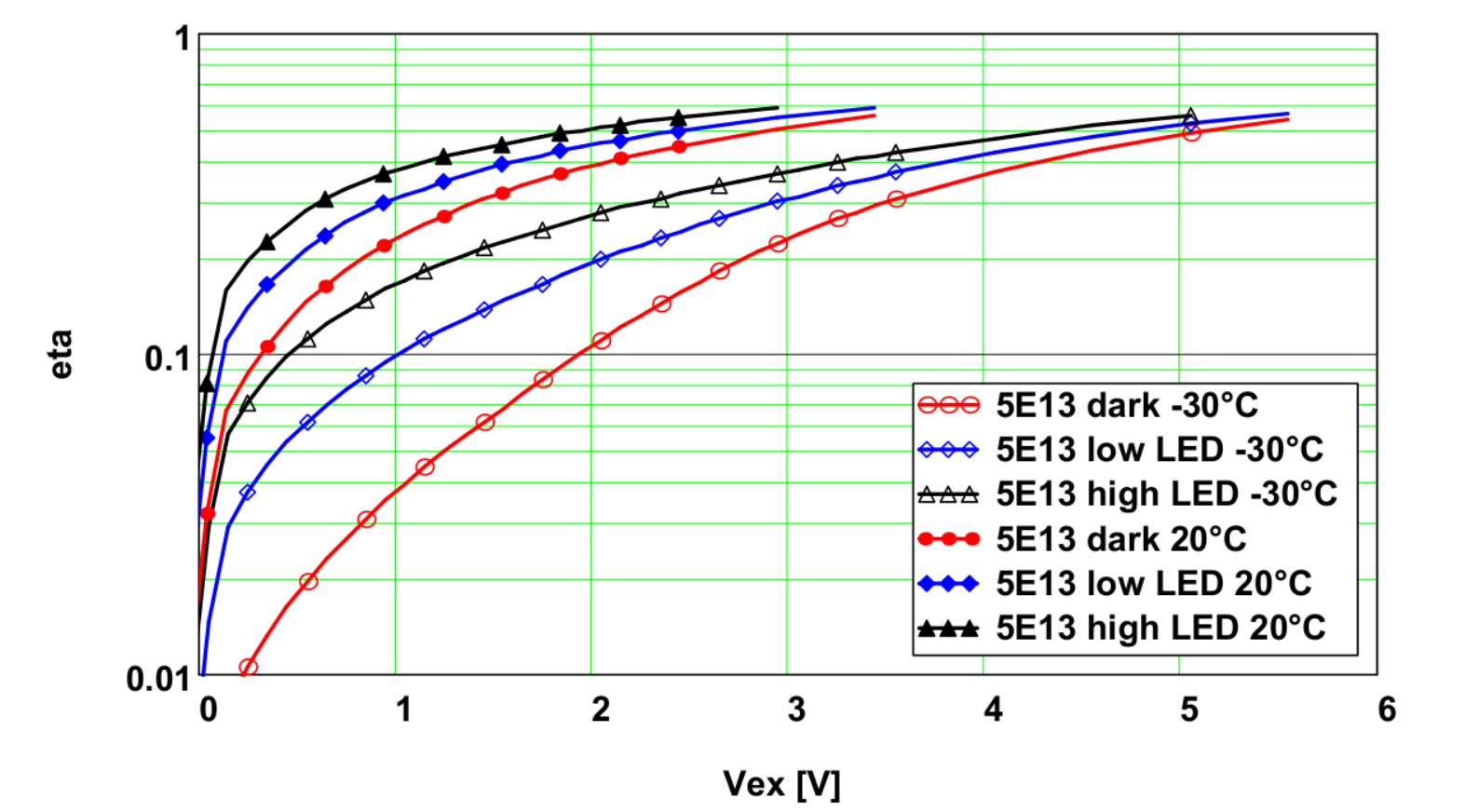}
    \caption{ }
   \end{subfigure}%
   \caption{Pixel occupancies, $\eta $ calculated using Eq.\,\ref{equ:eta} for (a) the non-irradiated, and (b) the SiPM irradiated to $5 \times 10^{13}$\,cm$^{-2}$. }
  \label{fig:eta}
 \end{figure}

 Fig.\,\ref{fig:eta} shows the pixel occupancies, $\eta _{DC}$, $\eta _{DC+lowLED}$ and $\eta _{DC+highLED}$ calculated using Eq.\,\ref{equ:eta}.
 For the non-irradiated SiPM (Fig.\,\ref{fig:eta}\,a), the $\eta$\,values are well below $10^{-3}$ for all measurement conditions, and no significant reduction of the $pde$ due to the occupancy is expected.
 For the SiPM irradiated to $5 \times 10^{13}$\,cm$^{-2}$ (Fig.\,\ref{fig:eta}\,b), the $\eta$\,values increase rapidly with $V_{ex}$ reaching values close to 60\,\%, and we expect a significant decrease in $pde$, as photons hitting a pixel in coincidence with a Geiger discharge from a dark count, will  produce no or reduced signals.
 As expected, $\eta _{DC+highLED} > \eta _{DC+lowLED} > \eta _{DC}$, and the increase of $\eta $ due to photons has to be taken into account in the analysis.

\begin{figure}[!ht]
   \centering
    \includegraphics[width=0.5\textwidth]{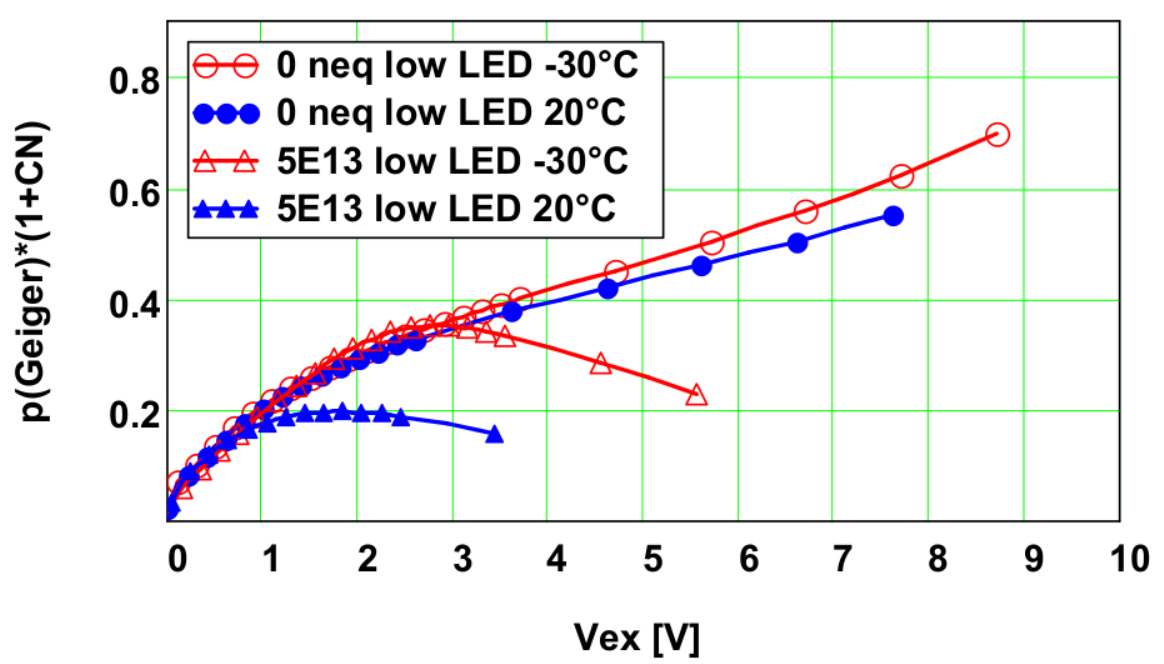}
   \caption{Product $p_{Geiger} \cdot (1 + CN)$, calculated using Eq.\,\ref{equ:I_LEDmodel}, as a function of $V_{ex}$. }
   \label{fig:pGeiger}
 \end{figure}

 Fig.\,\ref{fig:pGeiger}\, shows the product $p_{Geiger} \cdot (1 + CN)$ calculated using Eq.\,\ref{equ:I_LEDmodel} as a function of $V_{ex}$.
 For the non-irradiated SiPM the curve rises quickly above $V_{bd}$, and after some flattening continues to rise at higher values of $V_{ex}$.
 The rise for  $- 30 \,^\circ$C is faster than for $ 20 \,^\circ$C.
 Based on the results of Ref.\,\cite{Chmill:2017}, we ascribe this rise to the increase of correlated noise, $CN$, with $V_{ex}$.
 The $\Phi = 5 \times 10^{13}$\,cm$^{-2}$\,data follow the curves of the non-irradiated SiPM up to $\approx 0.75$\,V for the $ 20 \,^\circ$C\,data, and up to $\approx 2.5$\,V for the $ -30 \,^\circ$C\,data, and after reaching a maximum, decrease.
 We assume that this decrease is due to the high occupancy of the individual pixels: The pixels do not reach anymore the full biasing voltage.

   \subsection{Dark-count rate $DCR$}
  \label{sect:DCR-analysis}

\begin{figure}[!ht]
   \centering
    \includegraphics[width=0.5\textwidth]{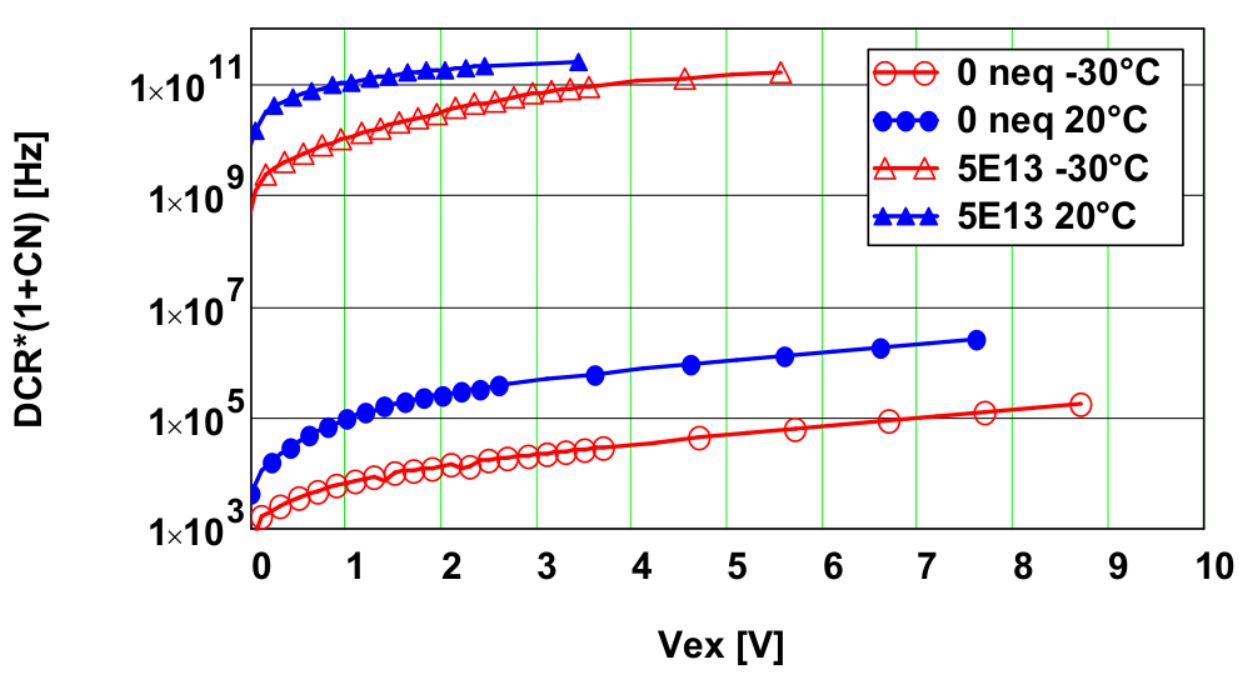}
   \caption{Product $DCR \cdot (1 + CN)$, calculated using Eq.\,\ref{equ:DCR}, as a function of $V_{ex}$. }
   \label{fig:DCR}
 \end{figure}

 Fig.\,\ref{fig:DCR} shows the product $DCR \cdot (1 + CN)$ calculated using Eq.\,\ref{equ:DCR} as a function of $V_{ex}$.
 As reported in Ref.\,\cite{Chmill:2017}, typical values of $CN$ at $20 \, ^\circ$C for the KETEK SiPM investigated are 0.05 at $V_{ex} = 3.5$\,V and 0.20 at 7.5\,V.
 It can be seen that the $DCR$ increases by approximately an order of magnitude between $-30 \,^\circ $C and $ +20 \,^\circ $C, and by about six orders of magnitude between no irradiation and irradiation to $5 \times 10^{13} $\,cm$^{-2}$.
 The value $DCR = 10^{11}$\,Hz corresponds to 10\,000 dark counts for a 100\,ns gate, which is typically used for pulse-height measurements.
 Thus, deriving the $DCR$ from pulse-height spectra under these conditions is difficult, if not impossible.

 \section{Summary and conclusions}
  \label{sect:Summary}

 In this note a fairly complete set of formulae is derived, which allows characterising SiPMs before and after irradiation with hadrons using current-voltage measurements with and without illumination.
 Central to the proposed method is the concept of the pixel occupancy, $\eta $, the probability that a pixel is \emph{busy} or \emph{occupied} by a Geiger discharge during a time interval $\Delta t$, from which the loss in photo-detection efficiency due to pile-up can be estimated.
 For $\Delta t$ the recharging time of a pixel $\tau = R_q \cdot C_{pix}$ is  assumed.
 The SiPM parameters determined with the proposed method are the quenching resistance, $R_q$, the breakdown voltage, $V_{bd}$, the pixel occupancy, $\eta$, the products of correlated noise times Geiger discharge probability, $(1 + CN) \cdot p_{Geiger} $, and dark-count rate, $(1 + CN) \cdot DCR $.
 The formulae allow to extend the characterisation of SiPMs into the region of $DCR$s well above $10 ^{10}$\,Hz, where other methods, like the analysis of pulse height spectra or transient current measurements, are having difficulties.
 Although the derivation of the formulae is straight-forward, the understanding of the validity of the assumptions made is much less obvious.

 To illustrate the application of the formulae and to investigate their validity, they are used to analyse current-voltage characteristics of a KETEK SiPM with 15\,$\upmu $m pixel size, measured with and without illumination by light from an LED at temperatures of $- 30 \, ^\circ$C and $+ 20 \, ^\circ$C, before and after neutron irradiation to a fluence of $5 \times 10^{13}$\,cm$^{-2}$, where dark-count rates exceeding $10^{11}$\,Hz are observed.

 Further applications of the formulae for the analysis of SiPM data will show to which extent they are valid, and which of the proposed methods are of practical use.
 The reader is strongly encouraged to try the formulae for the analysis of his SiPM measurements, and communicate failures and successes.
 This will help to improve the method.
 In addition, an attempt is made to clearly define all technical terms used.
 It is hoped that this will help to find a common nomenclature for the characterisation of SiPMs -- photo detectors with a most promising future.



 \input{bibliography}

  \label{sect:Bibliography}



\end{document}

%% file: bibliography.tex
%
%
\section{List of References}